\documentclass{ws-procs9x6}

\begin{document}

\title{Baryons, IN$_{\rm c}$}

\author{Richard F. Lebed}

\address{Department of Physics \& Astronomy, \\
Arizona State University, \\ 
Tempe, AZ 85287-1504, USA\\ 
E-mail: richard.lebed@asu.edu}

\maketitle

\abstracts{Excited baryons may be analyzed in the $1/N_c$ expansion
as true resonances in scattering amplitudes.  The key idea making this
program possible is a generalization of methods originally applied to
chiral soliton models in the 1980's.  One finds model-independent
relations among amplitudes that impose mass and width degeneracies
among resonances of various quantum numbers.  Phenomenological
evidence demonstrates that patterns of resonant decay predicted by
$1/N_c$ agree with data.  The analysis can be extended to subleading
orders in $1/N_c$, where again agreement with data is evident.}

\section{Introduction}

Although 100 of the 1000 pages in the {\it Review of Particle
Properties}\cite{PDG} consist of compiled measured properties of
baryon resonances, these states have resisted a model-independent
analysis for decades.  No one really understands {\it ab initio\/}
from QCD why baryon resonances exist at all, much less their peculiar
observed multiplicities, spacings, and branching fractions.  Even the
unambiguous existence of numerous resonances remains open to debate,
as evidenced by the infamous 1- to 4-star classification system.

Baryon resonances are exceptionally difficult to study precisely
because they {\em are\/} resonances, {\it i.e.}, unstable under strong
decay.  As a particular example, quark potential models are strictly
speaking applicable only when $q \bar q$ pair production from the
vacuum is suppressed, but this mechanism is the means by which baryon
resonances are produced in meson-baryon scattering.

The most natural description of excited baryons in large $N_c$ employs
the $N_c$ valence quark baryon picture: Since the ground-state baryon
multiplets ($J^P \! = \! {\frac 1 2}^+$ and ${\frac 3 2}^+$ for $N_c \! =
\! 3$) neatly fill a single multiplet completely symmetric under combined
spin-flavor symmetry [the SU(6) {\bf 56}], one may suppose that the
ground state of $O(N_c)$ quarks is also completely spin-flavor
symmetric.\footnote{This is reasonable because SU(6) spin-flavor
symmetry for ground-state baryons becomes exact in the large $N_c$
limit.\cite{DM}} Then, in analogy to the nuclear shell model, excited
states may be formed by promoting a small number [$O(N_c^0)$] of
quarks into orbitally or radially excited orbitals.  For example, the
generalization of the SU(6)$\times$O(3) multiplet $({\bf 70}, 1^- )$
consists of $N_c - 1$ quarks in the ground state and one in a relative
$\ell \! = \! 1$ state.  One may then analyze observables such as
masses and axial-vector couplings by constructing a Hamiltonian whose
terms possess definite transformation properties under the spin-flavor
symmetry and are accompanied by known powers of $N_c$.  By means of
the Wigner-Eckart theorem, one then relates observables for different
states in each multiplet.  This approach has been extensively
studied\cite{CGKM,Goity,PY,CCGL,GSS,CC} (see Ref.~\refcite{NStar} for
a short review), but it falls short in two important respects:

First, a Hamiltonian formalism is not entirely appropriate to unstable
particles, since it refers to matrix elements between asymptotic
external states.  Indeed, a resonance is properly represented by a
complex-valued pole in a scattering amplitude, for which the real and
imaginary parts indicate the mass and width, respectively.  Moreover,
the Hamiltonian approach makes no reference to resonances as
excitations of ground-state baryons.

Second, even if one were to construct a Hamiltonian respecting the
instability of the resonances, it is not clear that the simple
quark-shell baryon multiplets would be eigenstates of this
Hamiltonian.  Just as in the nuclear shell model, the possibility of
{\it configuration mixing\/} means that the true eigenstates might
consist of mixtures with 1, 2, or more excited quarks.

In contrast to quark potential models, chiral soliton models naturally
accommodate baryon resonances as excitations resulting from scattering
of mesons off ground-state baryons.  Such models are consistent with
the large $N_c$ limit because the solitons are heavy, semiclassical
objects compared to the mesons.  As has been known for many
years,\cite{ANW} a number of predictions following from the Skyrme and
other chiral soliton models are independent of the details of the
soliton structure, and may be interpreted as group-theoretical,
model-independent large $N_c$ results.  Indeed, the equivalence of
group-theoretical results for ground-state baryons in the Skyrme and
quark models in the large $N_c$ limit was demonstrated\cite{Manohar}
long ago.

In the remainder of this paper I discuss how the chiral soliton
picture may be used to study baryon resonances as well as the full
scattering amplitudes in which they appear.  It is based upon a series
of papers written in collaboration with Tom Cohen (and more recently
his students).\cite{CL1,CLcompat,CDLN1,CLpent,CDLN2}

\section{Amplitude Relations} \label{amp}

In the mid-1980's a series of papers\cite{HEHW,MK,MP,Mat3,MM}
uncovered a number of linear relations between meson-baryon scattering
amplitudes in chiral soliton models.  It became apparent that these
results are consistent with the large $N_c$ limit because of their
fundamentally group-theoretical nature.

Standard $N_c$ counting\cite{Witten} shows that ground-state baryons
have masses of $O(N_c^1)$, but meson-baryon scattering amplitudes are
$O(N_c^0)$.  Therefore, the characteristic resonant energy of
excitation above the ground state and resonance widths are both
generically expected to be $O(N_c^0)$.  To say that two baryon
resonances are degenerate to leading order in $1/N_c$ thus actually
means that their masses are equal at both the $O(N_c^1)$ and
$O(N_c^0)$ level.

An archetype of these linear relations was first derived in
Ref.~\refcite{MP}.  For a ground-state ($N$ or $\Delta$) baryon of
spin = isospin $R$ scattering with a meson (indicated by the
superscript) of relative orbital angular momentum $L$ (and primes for
analogous final-state quantum numbers) through a combined channel of
isospin $I$ and spin $J$, the full scattering amplitudes $S$ may be
expanded in terms of a smaller set of ``reduced'' scattering
amplitudes $s$:
\begin{eqnarray}
S_{LL^\prime R R^\prime IJ}^\pi & \! = & (-1)^{R^\prime \! \!
- R} \sqrt{[R][R^\prime]} \sum_K [K]
\left\{ \begin{array}{ccc} K &
I & J\\ R^\prime & L^\prime & 1 \end{array} \right\} \left\{
\begin{array}{ccc} K & I & J \\ R & L & 1 \end{array} \right\}
s_{K L^\prime L}^\pi \ , \ \label{MPeqn1} \\
S_{L R J}^\eta & = & \sum_K
\delta_{KL} \, \delta (L R J) \, s_{K}^\eta \ ,\label{MPeqn2}
\end{eqnarray}
where $[X] \equiv 2X+1$, and $\delta(j_1 j_2 j_3)$ indicates the
angular momentum triangle rule.\footnote{Both are consequences of a
more general formula\cite{Mat9j} involving $9j$ symbols that holds for
mesons of arbitrary spin and isospin, which for brevity we decline to
include.}  The fundamental feature inherited from chiral soliton
models is the quantum number $K$ ({\it grand spin}), with {\bf
K}$\,\equiv\,${\bf I}$\,$+$\,${\bf J}, conserved by the underlying
hedgehog configuration, which breaks $I$ and $J$ separately.  The
physical baryon state consists of a linear combination of $K$
eigenstates that is an eigenstate of both $I$ and $J$ but no longer
$K$.  $K$ is thus a good (albeit hidden) quantum number that labels
the reduced amplitudes $s$.  The dynamical content of relations such
as Eqs.~(\ref{MPeqn1})--(\ref{MPeqn2}) lies in the $s$ amplitudes,
which are independent for each value of $K$ allowed by $\delta (IJK)$.

In fact, $K$ conservation turns out to be equivalent to the $1/N_c$
limit.  The proof\cite{CL1} begins with the observation that the
leading-order (in $1/N_c$) $t$-channel exchanges have $I_t =
J_t,$\cite{KapSavMan} which in turn is proved using large $N_c$ {\em
consistency conditions}\cite{DJM}---essentially, unitarity
order-by-order in $N_c$ in meson-baryon scattering processes.
However, ($s$-channel) $K$ conservation was found---years earlier---to
be equivalent to the ($t$-channel) $I_t = J_t$ rule.\cite{MM} The
proof of this last statement relies on the famous Biedenharn-Elliott
sum rule,\cite{Edmonds} an SU(2) identity.

The significance of Eqs.~(\ref{MPeqn1})--(\ref{MPeqn2}) lies in the
fact that more full observable scattering amplitudes $S$ than reduced
amplitudes $s$ exist.  Therefore, one finds a number of linear
relations among the measured amplitudes holding at leading
[$O(N_c^0)$] order.  In particular, a resonant pole appearing in some
physical amplitude must appear in at least one reduced amplitude; but
this same amplitude contributes to a number of other physical
amplitudes, implying a degeneracy between the masses and widths of
resonances in several channels.\cite{CL1} For example, we apply
Eqs.~(\ref{MPeqn1})--(\ref{MPeqn2}) to negative-parity\footnote{Parity
enters by restricting allowed values of $L,L^\prime$.\cite{CLcompat}}
$I \! = \! \frac 1 2$, $J \! = \frac 1 2$ and $\frac 3 2$ states
(called $N_{1/2}$, $N_{3/2}$) in Table~\ref{I}.  Noting that neither
the orbital angular momenta $L , L^\prime$ nor the mesons $\pi , \eta$
that comprise the asymptotic states can affect the compound state
except by limiting available total quantum numbers ($I$, $J$, $K$),
one concludes that a resonance in the $S_{11}^{\pi NN}$ channel ($K \!
= \! 1$) implies a degenerate pole in $D_{13}^{\pi NN}$, because the
latter contains a $K \! = \!  1$ amplitude.
\renewcommand\arraystretch{1.25}%
\begin{table}[h]
\tbl{Application of Eqs.~(\ref{MPeqn1})--(\ref{MPeqn2}) to sample
negative-parity channels.
\label{I}}
{\footnotesize
\begin{tabular}{lcccccl}
\hline
State \mbox{ } && Quark Model Mass \mbox{ } &&
\multicolumn{3}{l}{Partial Wave, $K$-Amplitudes} \\
\hline
$N_{1/2}$ && $m_0$, $m_1$
   && $S^{\pi N N}_{11}$            &=& $s^\pi_{100}$ \\
&& && $D^{\pi \Delta \Delta}_{11}$  &=& $s^\pi_{122}$ \\
&& && $S^{\eta N N}_{1 1}$          &=& $s^\eta_0$ \\
\hline
%
%
$N_{3/2}$ && $m_1$, $m_2$
   && $D^{\pi N N}_{13}$            &=& $\frac 1 2
\left( s^\pi_{122} + s^\pi_{222} \right)$ \\
&& && $D_{13}^{\pi N \Delta}$       &=& $\frac 1 2
\left( s^\pi_{122} - s^\pi_{222} \right)$ \\
&& && $S_{13}^{\pi \Delta \Delta}$  &=& $s^\pi_{100}$ \\
&& && $D_{13}^{\pi \Delta \Delta}$  &=& $\frac 1 2
\left( s^\pi_{122} + s^\pi_{222} \right)$ \\
&& && $D_{13}^{\eta N N}$           &=& $s^\eta_2$ \\
\hline
%
%
%
%
\end{tabular} }
\end{table}
\renewcommand\arraystretch{1.0}%
One thus obtains towers of degenerate negative-parity resonance
multiplets labeled by $K$:
%
\begin{eqnarray}
N_{1/2} , \; \Delta_{3/2} , \; \cdots \; &~& (s_{0}^\eta) \; ,
\nonumber \\
N_{1/2} , \; \Delta_{1/2} , \; N_{3/2} , \; \Delta_{3/2} , \;
\Delta_{5/2} , \; \cdots \; &~& (s_{1 0 0}^\pi , s_{1 2 2}^\pi)
\; , \nonumber \\
\Delta_{1/2} , \; N_{3/2} , \; \Delta_{3/2} , \; N_{5/2} , \;
\Delta_{5/2} , \; \Delta_{7/2} , \; \cdots \;
&~& (s_{2 2 2}^\pi, s_{2}^\eta ) \; .
\label{towers}
\end{eqnarray}
%

It is now fruitful to consider the quark-shell large $N_c$ analogue of
the first excited negative-parity multiplet [the $({\bf 70}, 1^- )$].
Just as for $N_c \! = \! 3$, there are 2 $N_{1/2}$ and 2 $N_{3/2}$
states.  If one computes the masses to $O(N_c^0)$ for the entire
multiplet in which these states appear, one finds only three distinct
eigenvalues,\cite{CCGL,CL1,PS} which are labeled $m_0$, $m_1$, and
$m_2$ and listed in Table~\ref{I}.  Upon examining an analogous table
containing all the states in this multiplet,\cite{CL1} one quickly
concludes that exactly the required resonant poles are obtained if
each $K$ amplitude, $K \! = \! 0,1,2$, contains precisely one pole,
which located at the value $m_K$.  The lowest quark-shell multiplet of
negative-parity excited baryons is found to be {\em compatible\/}
with, {\it i.e.}, consist of a complete set of, multiplets classified
by $K$.

One can prove\cite{CLcompat} this compatibility for all nonstrange
multiplets of baryons in the SU(6)$\times$O(3) shell
picture.\footnote{Studies to extend these results to flavor SU(3) are
underway; while the group theory is more complicated, it remains
tractable.}  It is important to point out that compatibility does not
imply SU(6) is an exact symmetry at large $N_c$ for resonances as it
is for ground states.\cite{DM} Instead, it says that SU(6)$\times$O(3)
multiplets are {\em reducible\/} multiplets at large $N_c$.  In the
example given above, $m_{0,1,2}$ each lie only $O(N_c^0)$ above the
ground state, but are separated by amounts of $O(N_c^0)$.

We point out that large $N_c$ by itself does not mandate the existence
of any resonances at all; rather, it merely tells us that if even one
exists, it must be a member of a well-defined multiplet.  Although the
soliton and quark pictures both have well-defined large $N_c$ limits,
compatibility is a remarkable feature that combines them in a
particularly elegant fashion.

\section{Phenomenology} \label{phenom}

Confronting these formal large $N_c$ results with experiment poses two
significant challenges, both of which originate from neglecting
$O(1/N_c)$ corrections.  First, the lowest multiplet of nonstrange
negative-parity states covers quite a small mass range (only
1535--1700~MeV), while $O(1/N_c)$ mass splittings can generically be
as large as $O$(100~MeV).  Any claims that two such states are
degenerate while two others are not must be carefully scrutinized.
Second, the number of states in each multiplet increases with $N_c$,
meaning that a number of large $N_c$ states are spurious in $N_c \! =
\! 3$ phenomenology.  For example, for $N_c \! \ge \! 7$ the analogue
of the {\bf 70} contains 3 $\Delta_{3/2}$ states, but only 1
[$\Delta(1700)$] for $N_c \! = \! 3$.  As $N_c$ is tuned down from
large values toward 3, the spurious states must decouple through the
appearance of factors such as $(1-3/N_c)$, which in turn requires one
to understand simultaneously leading and subleading terms in the
$1/N_c$ expansion.

Nevertheless, it is possible to obtain testable predictions for the
decay channels, even with just the leading-order results.  For
example, note from Table~\ref{I} that the $K \! = \! 0(1)$ $N_{1/2}$
resonance couples only to $\eta$($\pi$).  Indeed, despite lying barely
above the $\eta N$ threshold, the $N(1535)$ resonance decays through
this channel 30--55$\%$ of the time, while the $N(1650)$, which has
much more comparable phase space for $\pi N$ and $\eta N$, decays to
$\eta N$ only 3--10$\%$ of the time.  This pattern clearly suggests
that the $\pi$-phobic $N(1535)$ should be identified with $K \! = \!
0$ and the $\eta$-phobic $N(1650)$ with $K \! = \! 1$, the first fully
field theory-based explanation for these phenomenological facts.

\section{Configuration Mixing}

As mentioned above, quark-shell baryon states with a fixed number of
excited quarks are not expected to be eigenstates of the full QCD
Hamiltonian.  Rather, configuration mixing is expected to cloud the
situation.  Consider, for example, the statement that baryon
resonances are expected to have generically broad [$O(N_c^0)$] widths.
One may ask whether some states might escape this restriction and turn
out to be narrow in the large $N_c$ limit.  Indeed, some of the first
work\cite{PY} on excited baryons combined large $N_c$ consistency
conditions and a quark description of excited baryon states to predict
that baryons in the {\bf 70}-analogue have widths of $O(1/N_c)$, while
states in an excited negative-parity spin-flavor symmetric multiplet
({\bf 56$^\prime$}) have $O(N_c^0)$ widths.

In fact there exist, even in the quark-shell picture, operators
responsible for configuration mixing between these
multiplets.\cite{CDLN1} The spin-orbit and spin-flavor tensor
operators (respectively $\ell s$ and $\ell^{(2)} g \, G_c$ in the
notation of Refs.~\refcite{CCGL},$\,$\refcite{GSS},$\,$\refcite{PS}),
which appear at $O(N_c^0)$ and are responsible for splitting the
eigenvalues $m_0$, $m_1$, and $m_2$, give nonvanishing transition
matrix elements between the {\bf 70} and {\bf 56$^\prime$}.  Since
states in the latter multiplet are broad, configuration mixing forces
at least some states in the former multiplet to be broad as well.  One
concludes that the possible existence of any excited baryon state
narrow in the large $N_c$ limit requires a fortuitous absence of
significant configuration mixing.

\section{Pentaquarks}

The possible existence of a narrow isosinglet, strangeness +1 (and
therefore exotic) baryon state $\Theta^+ (1540)$, claimed to be
observed by numerous experimental groups (but not seen by several
others), was much discussed at this meeting.  Although the jury
remains out on this important question, one may nevertheless use the
large $N_c$ methods described above to determine what degenerate
partners a state with these quantum numbers would
possess.\cite{CLpent} For example, if one imposes the theoretical
prejudice $J_{\Theta} \! = \! \frac 1 2$, then there must also be
pentaquark states with $I \! = \! 1$, $J \! = \! \frac 1 2, \frac 3 2$
and $I \! = \! 2$, $J \! = \! \frac 3 2, \frac 5 2$, with masses and
widths equal that of the $\Theta^+$, up to $O(1/N_c)$ corrections.

The large $N_c$ analogue of the ``pentaquark'' actually carries the
quantum numbers of $N_c \! + \! 2$ quarks, consisting of $(N_c \!  +
\! 1)/2$ spin-singlet, isosinglet $ud$ pairs and an $\bar s$ quark.
The operator picture, for example, shows the partner states we predict
to belong to SU(3) multiplets {\bf 27} ($I \! = \! 1$) and {\bf 35}
($I \! = \! 2$).\cite{JMpent} However, the existence of partners does
not depend upon any particular picture for the resonance or any
assumptions regarding configuration mixing.  Since the generic width
for such baryon resonances remains $O(N_c^0)$, the surprisingly small
reported width ($<$10 MeV) does not appear to be explicable by large
$N_c$ considerations alone, but may be a convergence of small phase
space and a small nonexotic-exotic-pion coupling.

\section{$1/N_c$ Corrections}

All the results exhibited thus far hold at the leading nontrivial
order ($N_c^0$) in the $1/N_c$ expansion.  We saw in Sec.~\ref{phenom}
that $1/N_c$ corrections are essential not only to explain the sizes
of effects apparent in the data, but in the very enumeration of
physical states.  Clearly, if this analysis is to carry real
phenomenological weight, one must demonstrate a clear path to
characterize $1/N_c$ corrections to the scattering amplitudes.
Fortunately, such a generalization is possible: As discussed in
Sec.~\ref{amp}, the constraints on scattering amplitudes obtained from
the large $N_c$ limit are equivalent to the $t$-channel requirement
$I_t \! = \! J_t$.  In fact, Refs.~\refcite{KapSavMan} showed not only
that this result holds in the large $N_c$ limit, but also that
exchanges with $|I_t \! - \! J_t| \! = \! n$ are suppressed by a
relative factor $1/N_c^n$.

This result permits one to obtain relations for the scattering
amplitudes including all effects up to and including $O(1/N_c)$:
\begin{eqnarray}
\! \! \! \! S_{LL^\prime R R^\prime I_s J_s} \! \! \! \! & = &
\sum_{\mathcal J} \left[
\begin{array}{ccc} 1 & R^\prime & I_s \\ R & 1 & {\mathcal J}
\! = \! I_t
\end{array} \right] \left[
\begin{array}{ccc} L^\prime & R^\prime & J_s \\ R & L & {\mathcal J}
\! = \! J_t
\end{array}
\right] s_{{\mathcal J} L L^\prime}^t \nonumber \\ &
-\frac{1}{N_c} & \sum_{\mathcal J} \left[ \begin{array}{ccc} 1 &
R^\prime & I_s\\ R & 1 & I_t \! = \! {\mathcal J}
\end{array} \right] \left[
\begin{array}{ccc} L^\prime & R^\prime & J_s \\ R & L &
J_t \! = \! {{\mathcal J} \! + \! 1}
\end{array}
\right] s_{{\mathcal J} L L^\prime}^{t(+)} \nonumber \\ &
- \frac{1}{N_c} & \sum_{\mathcal J} \left[
\begin{array}{ccc} 1 & R^\prime & I_s\\ R & 1 & I_t \! = \!
{\mathcal J}
\end{array} \right] \left[
\begin{array}{ccc} L^\prime & R^\prime & J_s \\ R & L &
J_t \! = \! {{\mathcal J} \! - \! 1}
\end{array}
\right] s_{{\mathcal J} L L^\prime}^{t(-)} + 
O(\mbox{\small{$\frac{1}{N_c^2}$}}) ,
\label{MPplus}
\end{eqnarray}
One obtains this expression by first rewriting $s$-channel expressions
such as Eqs.~(\ref{MPeqn1})--(\ref{MPeqn2}) in terms of $t$-channel
amplitudes.  The $6j$ symbols in this case contain $I_t$ and $J_t$ as
arguments (which for the leading term are equal).  One then
introduces\cite{CDLN2} new $O(1/N_c)$-suppressed amplitudes
$s^{t(\pm)}$ for which $J_t - I_t = \pm 1$.  The square-bracketed $6j$
symbols in Eq.~(\ref{MPplus}) differ from the usual ones only through
normalization factors, and in particular obey the same triangle rules.

\begin{figure}[t]
\centerline{\epsfxsize=4.1in\epsfbox{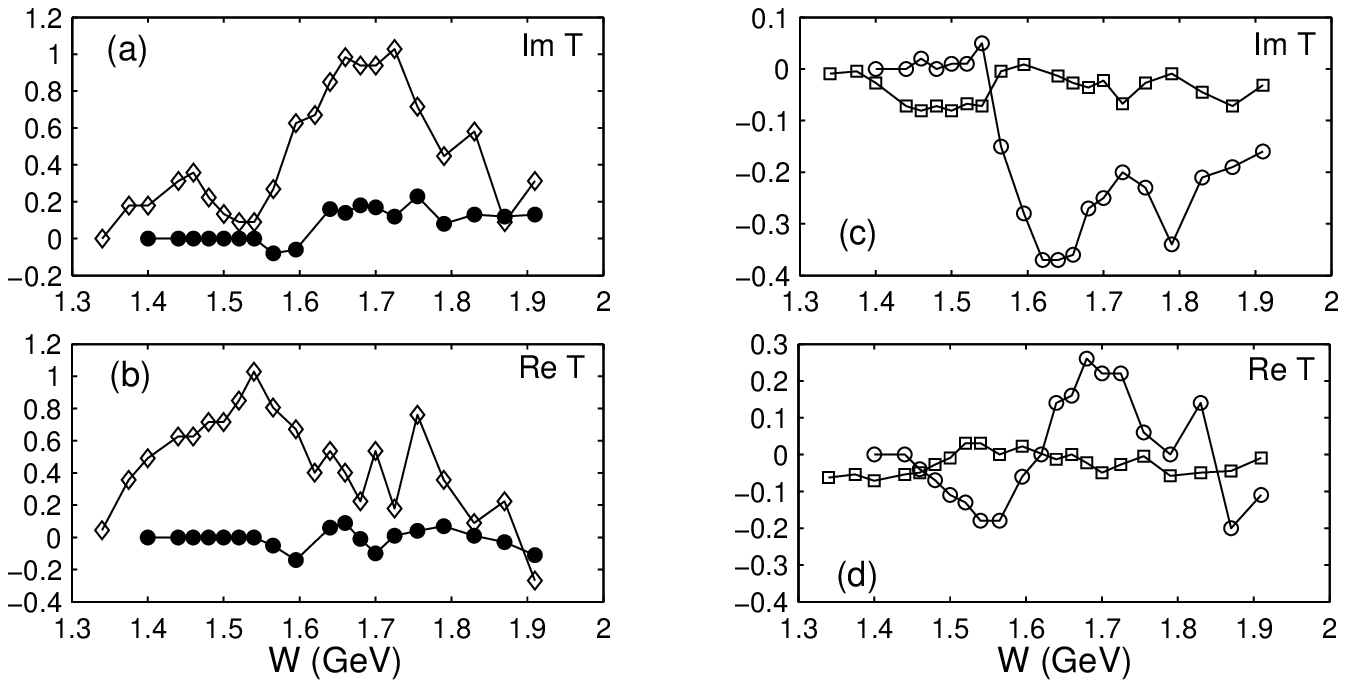}}
\centerline{\epsfxsize=4.1in\epsfbox{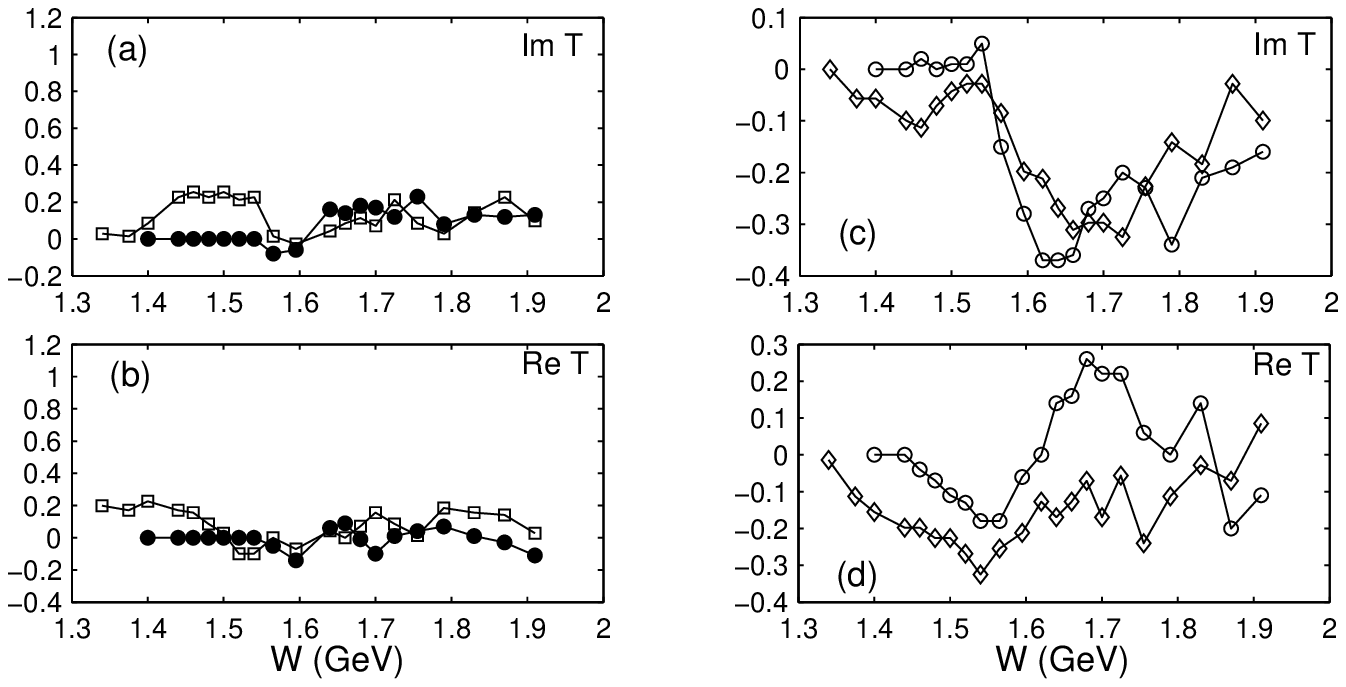}}   
\caption{Real and imaginary parts of $\pi N \to \pi \Delta$ scattering
amplitudes.  The first 4 insets give two particular partial waves
equal to leading order [hence indicating the size of $O(1/N_c)$
corrections].  The second 4 insets give two particular linear
combinations of the same data good to $O(1/N_c^2).$\label{inter}}
\end{figure}

Relations between observable amplitudes that incorporate the larger
set $s^t$, $s^{t(+)}$, and $s^{t(-)}$ are expected to hold a factor of
3 better than those merely including the leading $O(N_c^0)$ results.
Indeed, this is dramatically evident in cases where sufficient data is
available, $\pi N \to \pi \Delta$ scattering (Fig.~\ref{inter}).  For
example, (c) and (d) in the first 4 insets give the imaginary and real
parts, respectively, of partial wave data for the channels $SD_{31}$
({\Large\rm$\circ$}) and $(1/\sqrt{5}) DS_{13}$ ($\square$), which are
equal up to $O(1/N_c)$ corrections; in (c) and (d) of the second 4
insets, the {\Large\rm $\circ$} points again are $SD_{31}$ data, while
$\lozenge$ represent $-\sqrt{2} DS_{33}$, and by Eq.~(\ref{MPplus})
these are equal up to $O(1/N_c^2)$ corrections.

\section{Conclusions}

The purpose of this talk has been to convince you that there now exist
reliable calculational techniques to handle not only long-lived
ground-state baryons, but also unstable baryon resonances and the
scattering amplitudes in which they appear.  This approach, originally
noted in chiral soliton models but eventually shown to be a true
consequence of large $N_c$ QCD, is found to have phenomenological
consequences [such as the large $\eta$ branching fraction of the
$N(1535)$] that compare favorably with real data.

The first few steps into studying and classifying $1/N_c$ corrections
to the leading-order results, absolutely essential to obtaining
comparison with the full data set, have begun.  The measured
scattering amplitudes appear to obey the constraints placed by these
corrections, and more work along these lines will be forthcoming.  For
example, the means by which the spurious extra resonances of large
$N_c$ decouple as one tunes the value of $N_c$ down to 3 is crucial
and not yet understood.

All the results presented here, as mentioned in Sec.~\ref{amp}, have
used only relations among states of fixed strangeness.  Moving beyond
this limitation means using flavor SU(3) group theory, which is rather
more complicated than isospin SU(2) group theory.  Nevertheless, this
is merely a technical complication, and existing work shows that it
can be overcome.\cite{GSS}

At the time of this writing, all of the necessary tools appear to be
in place to commence a full-scale analysis of baryon scattering and
resonance parameters.  One may envision a sort of resonance
calculation factory, {\em Baryons $IN_C$}.  With sufficient time and
researchers, the whole baryon resonance spectrum can be disentangled
using a solid, field-theoretical approach based upon a well-defined
limit of QCD.

\section*{Acknowledgments}

I would like to thank the organizers for their hospitality and
inviting me to this most lively conference.  The work described here
was supported in part by the National Science Foundation under Grant
No.\ PHY-0140362.


\begin{thebibliography}{99}

\bibitem{PDG}
{\it Review of Particle Properties} (K.~Hagiwara {\it et al.}), {\it
Phys.\ Rev.} {\bf D66}, 010001 (2002).

\bibitem{DM}
R.F.~Dashen and A.V.~Manohar, {\it Phys.\ Lett.} {\bf B315}, 425
(1993).

\bibitem{CGKM}
C.D.~Carone, H.~Georgi, L.~Kaplan, and D.~Morin, {\it Phys.\ Rev.}
{\bf D50}, 5793 (1994).

\bibitem{Goity}
J.L.~Goity, {\it Phys.\ Lett.} {\bf B414}, 140 (1997).

\bibitem{PY}
D.~Pirjol and T.-M.~Yan, {\it Phys.\ Rev.} {\bf D57}, 1449 (1998);
{\bf D57}, 5434 (1998).

\bibitem{CCGL}
C.E.~Carlson, C.D.~Carone, J.L.~Goity, and R.F.~Lebed, {\it Phys.\
Lett.} {\bf B438}, 327 (1998); {\it Phys.\ Rev.} {\bf D59}, 114008
(1999).

\bibitem{GSS}
J.L.~Goity, C.~Schat, and N.~Scoccola, Phys.\ Rev.\ Lett.\ {\bf 88},
102002 (2002); Phys.\ Rev.\ {\bf D66}, 114014 (2002); {\it Phys.\
Lett.} {\bf B564}, 83 (2003).

\bibitem{CC}
C.E.~Carlson and C.D.~Carone, Phys.\ Rev.\ {\bf D58}, 053005 (1998);
Phys.\ Lett.\ {\bf B441}, 363 (1998); {\bf B484}, 260 (2000).

\bibitem{NStar}
R.F.~Lebed, in {\it NStar 2002: Proceedings of the Workshop on the
Physics of Excited Nucleons}, ed.\ by S.A.~Dytman and E.S.~Swanson,
World Scientific, Singapore, 2003, p.~73 [{\tt hep-ph/0301279}].

\bibitem{ANW}
E.~Witten, {\it Nucl.\ Phys.} {\bf B223}, 433 (1983); G.S.~Adkins,
C.R.~Nappi, and E.~Witten, {\it Nucl.\ Phys.} {\bf B228}, 552 (1983);
G.S.~Adkins and C.R.~Nappi, {\it Nucl.\ Phys.} {\bf B249}, 507 (1985).

\bibitem{Manohar}
A.V.~Manohar, {\it Nucl.\ Phys.} {\bf B248}, 19 (1984).

\bibitem{CL1}
T.D.~Cohen and R.F.~Lebed, {\it Phys.\ Rev.\ Lett.} {\bf 91}, 012001
(2003); {\it Phys.\ Rev.} {\bf D67}, 012001 (2003).

\bibitem{CLcompat}
T.D.~Cohen and R.F.~Lebed, {\it Phys.\ Rev.} {\bf D68}, 056003 (2003).

\bibitem{CDLN1}
T.D.~Cohen, D.C.~Dakin, A.~Nellore, and R.F.~Lebed, {\it Phys.\ Rev.}
{\bf D69}, 056001 (2004).

\bibitem{CLpent}
T.D.~Cohen and R.F.~Lebed, {\it Phys.\ Lett.} {\bf B578}, 150 (2004).

\bibitem{CDLN2}
T.D.~Cohen, D.C.~Dakin, A.~Nellore, and R.F.~Lebed, {\tt
hep-ph/0403125}.

\bibitem{HEHW} 
A.~Hayashi, G.~Eckart, G.~Holzwarth, H.~Walliser, {\it Phys.\ Lett.}
{\bf B147}, 5 (1984).

\bibitem{MK}
M.P.~Mattis and M.~Karliner, {\it Phys.\ Rev.} {\bf D31}, 2833 (1985).

\bibitem{MP}
M.P.~Mattis and M.E.~Peskin, {\it Phys.\ Rev.} {\bf D32}, 58 (1985).

\bibitem{Mat3}
M.P.~Mattis, {\it Phys.\ Rev.\ Lett.} {\bf 56}, 1103 (1986); {\it
Phys.\ Rev.}  {\bf D39}, 994 (1989); {\it Phys.\ Rev.\ Lett.} {\bf
63}, 1455 (1989).

\bibitem{MM}
M.P.~Mattis and M.~Mukerjee, {\it Phys.\ Rev.\ Lett.} {\bf 61}, 1344
(1988).

\bibitem{Witten}
E.~Witten, {\it Nucl.\ Phys.} {\bf B160}, 57 (1979).

\bibitem{Mat9j}
M.P.~Mattis, {\it Phys.\ Rev.\ Lett.} {\bf 56}, 1103 (1986).

\bibitem{KapSavMan}
D.B.~Kaplan and M.J.~Savage, {\it Phys.\ Lett.} {\bf B365}, 244
(1996); D.B.~Kaplan and A.V.~Manohar, {\it Phys.\ Rev.} {\bf C56}, 76
(1997).

\bibitem{DJM}
R.F.~Dashen, E. Jenkins, and A.V.~Manohar, {\it Phys.\ Rev.} {\bf
D49}, 4713 (1994).

\bibitem{Edmonds}
A.R.~Edmonds, {\it Angular Momentum in Quantum Mechanics} (Princeton
Univ.\ Press, Princeton, NJ, 1996) [Eq.~(6.2.12)].

\bibitem{PS}
D.~Pirjol and C.~Schat, {\it Phys.\ Rev.} {\bf D67}, 096009 (2003).

\bibitem{JMpent}
E.~Jenkins and A.V.~Manohar, {\tt hep-ph/0401190} and {\tt 0402024}.

\end{thebibliography}
\end{document}